\documentclass[lettersize,journal]{IEEEtran}
\usepackage{hyperref}
\usepackage{url}
\usepackage{tablefootnote}
\usepackage[noadjust]{cite}
\usepackage{graphicx}
\usepackage{textcomp}
\usepackage{subcaption}
\usepackage{adjustbox}
\usepackage{caption}
\usepackage{makecell}
\usepackage{multirow}
\usepackage{xcolor}
\usepackage{threeparttable}
\usepackage{verbatim}
\usepackage{mathrsfs}
\usepackage{siunitx}
\usepackage{colortbl}
\usepackage{bm}
\usepackage{tablefootnote}
\usepackage{enumitem}
\usepackage{booktabs}
\usepackage{amsmath}
\usepackage{algorithm}
\usepackage{algorithmic}

\def\BibTeX{{\rm B\kern-.05em{\sc i\kern-.025em b}\kern-.08em
    T\kern-.1667em\lower.7ex\hbox{E}\kern-.125emX}}
\usepackage{balance}
\begin{document}
\title{A Wearable Strain-Sensor-Based Shoulder Patch for Fatigue Detection in Bicep Curls}
\author{Ming Xuan Chua,~\IEEEmembership{Student Member,~IEEE}, Shuhua Peng, Thanh Nho Do, Chun Hui Wang, and~Liao~Wu,~\IEEEmembership{Member,~IEEE}
\thanks{This work involved human subjects. Approval of the ethical, experimental procedures and protocols was granted by The University of New South Wales under the HC reference number: HC220573.} 
\thanks{Corresponding author: Liao Wu}
\thanks{Ming Xuan Chua, Shuhua Peng, Chun Hui Wang, and Liao Wu are with the School of Mechanical and Manufacturing Engineering, The University of New South Wales, Australia (email: mingxuan.chua@unsw.edu.au; shuhua.peng@unsw.edu.au; chun.h.wang@unsw.edu.au; dr.liao.wu@ieee.org).}
\thanks{Thanh Nho Do is with the Graduate School of Biomedical Engineering, The University of New South Wales, Australia (email: tn.do@unsw.edu.au).}
}
\markboth{IEEE TRANSACTIONS ON INSTRUMENTATION AND MEASUREMENT}%
{A Wearable Strain-Sensor-Based Shoulder Patch for Fatigue Detection in Bicep Curls}

\maketitle

\begin{abstract}
A common challenge in home-based rehabilitation is muscle compensation induced by pain or fatigue, where patients with weakened primary muscles recruit secondary muscle groups to assist their movement, causing issues such as delayed rehabilitation progress or risk of further injury. In a home-based setting, the subtle compensatory actions may not be perceived since physiotherapists cannot directly observe patients. 
To address this problem, this study develops a novel wearable strain-sensor-based shoulder patch to detect fatigue-induced muscle compensation during bicep curl exercises. 
Built on an observation that the amplitude of a strain sensor's resistance is correlated to the motion of a joint that the sensor is attached to, we develop an algorithm that can robustly detect the state when significant changes appear in the shoulder joint motion, which indicates fatigue-induced muscle compensation in bicep curls.
The developed shoulder patch is tested on 13 subjects who perform bicep curl exercises with a 5 kg dumbbell until reaching fatigue. 
During the experiment, the performance of the shoulder patch is also benchmarked with optical tracking sensors and surface electromyography (sEMG) sensors. Results reveal that the proposed wearable sensor and detection methods effectively monitor fatigue-induced muscle compensation during bicep curl exercises in both Real-Time and Post Hoc modes. This development marks a significant step toward enhancing the effectiveness of home-based rehabilitation by providing physiotherapists with a tool to monitor and adjust treatment plans remotely.

\end{abstract}


\section{Introduction}

Traditional rehabilitation typically requires one-on-one interactions with a physiotherapist to visually monitor the patients so that prescribed exercises are performed correctly. This approach faces several challenges, including high patient-to-physiotherapist ratios, the time and effort it takes for patients to access clinics, and the constraints imposed by appointment scheduling \cite{abedi2024artificial}. These issues are exacerbated in rural areas where limited availability of clinics further restricts patient access to these services, placing a substantial burden on the healthcare system and often resulting in suboptimal rehabilitation services \cite{manjunatha2021upper}. 

Home-based rehabilitation, conducted in the patient's residence rather than a clinical setting, can reduce these burdens but introduces its own challenges. A key concern is the lack of direct supervision by physiotherapists, making it difficult to monitor patient progress accurately. A crucial part of effectively monitoring patient progress in home-based rehabilitation is the ability to gauge patient’s effort. Substantial effort can quickly lead to fatigue, and assessing it has been challenging. In clinical rehabilitation, physiotherapists employ various methods, including subjective measures such as Borg Rate of Perceived Exertion (RPE) \cite{williams2017borg}, and objective measurements via physiological signal monitoring such as heart rate, EMG signals, and performance-based tests such as range of motion, force, speed, and completion of a task. However, fatigue is multifaceted, complex, and manifests differently across individuals. As a result, assessing fatigue is challenging and requires instinct from the healthcare provider.  

With fatigue, patients may unknowingly compensate during exercise, a behavior where the secondary muscles or joints are increasingly engaged to complete a movement when the primary muscle is at reduced capacity. For example, Liu et al. \cite{liu2021muscle} utilized complex networks theory, showing the shifts of the muscles' functional connection as fatigue increases, exemplifying the compensation mechanism. Although compensation helps patients accomplish the movement, it can lead to chronic issues such as pain, reduced range of motion, and diminished capacity for motor function in the affected joints \cite{levin2009motor}.

Our previous study investigated fatigue-induced muscle compensation in bicep curl exercises \cite{chua2024analysis}. We recruited 12 healthy male subjects and collected two key signals while they performed bicep curl exercises under non-fatigue and fatigue conditions: (1) the surface electromyography (sEMG) signals from eight different muscles, and (2) the upper limb joint kinematics. Through muscle synergy and joint kinematics analysis, we observed significant compensation during fatigue:

\begin{itemize}
    \item The relative contribution of the Upper Trapezius (UT) muscle increased by approximately 62.5\%, and the RMS amplitude of its sEMG signal increased by 127\%. 
    \item The shoulder joint's movement also intensified, including increased amplitude in both shoulder elevation-depression motion and shoulder flexion-extension motion. Notably, the shoulder elevation-depression range of motion (ROM) increased by about 150\%.
\end{itemize}
 
Based on these observations, our previous study suggested two effective methods to capture fatigue during bicep curl exercises, including (1) monitoring the sEMG signal of the UT muscle and (2) tracking the ROM of the shoulder joint. In both methods, the amplitude of the signal concerned is expected to increase significantly during fatigue compared to the non-fatigue state.
 
In \cite{chua2024analysis}, sEMG signals were collected from several sEMG sensors attached to the surface of the investigated muscles. These sensors measure the electrical signal generated by the muscle fibers during muscle contraction. The joint kinematics were recorded by a motion capture system, which utilized multiple infrared cameras positioned around the capture area to detect the reflective markers attached to the key points of the object to precisely track their movement.
Despite high accuracy, using sEMG sensors or motion capture systems in home-based rehabilitation is impractical due to their cost, complexity, and the need for trained professionals to operate them \cite{wang2022technology}. As a result, they have mainly been used to set up the ground truth for an exercise \cite{liu2022joint,sgambato2023high} or as a validation tool of a developed sensor system \cite{hadjipanayi2024remote,alvarez2022towards}. This highlights the necessity for more user-friendly alternatives that can capture muscle force and joint movements independently. 
For example, sensors such as Inertial Measurement Units (IMUs), which capture the acceleration and angular velocity \cite{hua2020evaluation,ding2024system}, pressure sensors, which capture the exerted force and the applied pressure \cite{cai2019automatic}, or marker-free cameras, which capture the visual information of movements \cite{tannous2019gamerehab,xu2022multiview}, have been maturely used for joint movement detection, particularly in rehabilitation settings.

However, they are susceptible to environmental influences such as sensor orientation and lighting conditions. For muscle force measurement, sensing options are more limited. Although mechanomyography (MMG) \cite{hondo2022torque}, which measures the mechanical oscillations of muscles during contractions, provides a direct assessment, it is less commonly used compared to sEMG sensors due to less technological maturity.
Dynamometers \cite{alvarez2022towards}, while precise, are typically confined to clinical settings due to their operational complexities. In addition, it only measures the torque generated during muscle contraction against the dynamometer resistance and does not directly measure muscle force. Therefore, developing an integrated device that combines these technologies into a robust, easy-to-use system for home rehabilitation is essential. Further research is needed to enhance the accuracy, usability, and environmental resilience of these sensors, making them suitable for non-professional users in home settings.

Strain gauge sensors have demonstrated versatility across various applications, including monitoring breathing rates, muscle force, and joint kinematics. Jonathan et al. \cite{alvarez2022towards} employed wearable soft strain sensors to track muscle deformation during contractions, discovering patterns of muscle activity similar to those detected by dynamometers. Sun et al. \cite{sun2022ultrasensitive} developed a strain sensor capable of capturing joint movements, such as the bending of fingers and wrists, further illustrating the sensor's utility in dynamic motion tracking. Tang et al. \cite{tang2024ultrasensitive} extended the application of strain sensors to develop a choker as an interface for silent speech, employing a neural network model to analyze strain sensor signals generated during silent speech attempts. Lin et al. \cite{lin2020jacket} crafted an ``E-Jacket" equipped with strain sensors to monitor different postures, using a CNN-LSTM model to analyze inputs from four sensors and classify user postures including standing, walking, sitting, running, and lying.

These examples highlight the broad potential of strain sensors, particularly in targeted applications for muscle activity detection and joint movement monitoring. However, these works mostly rely on arrays of strain sensors to determine joint movements via activated signals from a subset of sensors and use machine learning algorithms for complex tasks. Compared to others, strain sensors offer the advantage of providing direct measurements of the sensor formation on the skin surface, which often correlates closely with the underlying joint movement. Additionally, the strain sensors are more user-friendly due to their lightweight, conformability to the skin surface, and low setup required for effective use. 

Co-developing medical sensors employing machine learning methods typically requires extensive pre-trained data, necessitating numerous medical trials using the developed sensor, which often involve regulatory hurdles and variability in the patients' data, thereby increasing development costs and time. Consequently, there is a significant need to create a low-cost strain sensor system for fatigue detection, simplifying the technology and reducing reliance on extensive data sets for effective application in real-world settings.

To this end, this paper introduces a wearable strain-sensor-based shoulder patch to capture the onset of fatigue during bicep curl exercises. Building on the stretchable strain sensors developed by our team previously \cite{peng2021carbon}, we designed a shoulder patch that can be conveniently attached to a user for muscle activity monitoring. The patch is attached to the Acromioclavicular (AC) joint through kinesiology tape, and the shoulder movement is monitored by reading the strain sensor resistance. The obtained signal is processed in both Real-Time and Post Hoc modes. The former enables practical application in detecting a patient's state while exercising. The later allows for the retrospective analysis of patients' rehabilitation states to provide personalized rehabilitation feedback to the patient and understand their fatigue patterns.

To validate the effectiveness of this shoulder patch and its associated algorithms for fatigue detection, we recruited 13 subjects to perform bicep curl exercises with a 5kg dumbbell until fatigue, followed by additional bicep curl exercises under fatigue conditions. Throughout the study, the strain sensor signal, the sEMG signal of the UT muscle, and the shoulder elevation-depression trajectory were collected and analyzed to estimate the subjects' fatigue onset. The estimated fatigue times were then compared with the subjects' self-declared fatigue time, as well as the fatigue times derived from sEMG signals and shoulder kinematics. 

The key contributions of this study include:
\begin{itemize}
    \item Developing a wearable strain-sensor-based shoulder patch capable of detecting fatigue during bicep curl exercise;
    \item Proposing cost-effective algorithms to process the sensor signal in both Real-Time and Post Hoc modes, enhancing the practical utility of the sensor in fatigue detection; 
    \item Comparing the performance of the developed device and algorithms against sEMG and joint kinematics - two methods commonly used in lab experiments - and demonstrating the effectiveness of the proposed system for fatigue detection in home-based bicep curl rehabilitation.
\end{itemize}
The subsequent sections of this paper are organised as follows: Section \ref{sec:development} outlines the development of the proposed sensor system; Section \ref{sec:Ex&Valid} describes the experimental and validation protocols of the study; Section \ref{sec:result} presents the experimental results and discusses the comparison with the onset of fatigue from different sources. Finally, section \ref{sec:Conclusion} concludes the study and depicts the potential future work. 

\begin{figure*}[h!]	
    \centering
    \includegraphics[width=0.9\textwidth]{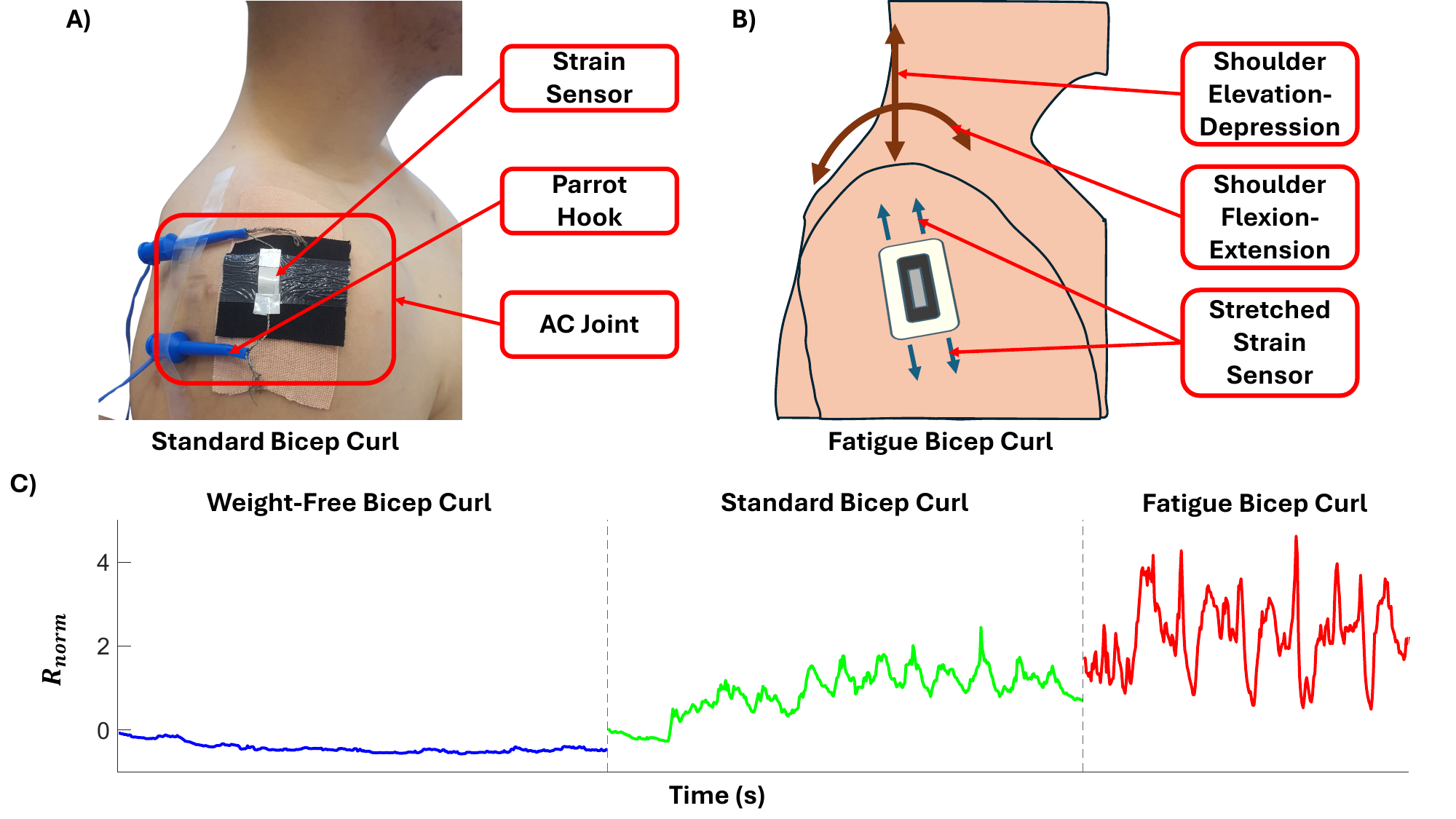}
    \caption{Overview of the study, employing the strain sensor for monitoring fatigue during bicep curl exercise. \textcolor{blue}{(A)} Illustration of the placement of the strain sensor on the subject during the exercise. The sensor is selected to be placed on the AC joint. The rationale of the sensor placement is discussed in Sec. \ref{subsec:Ex2}. \textcolor{blue}{(B)} The working principle of the sensor developed, designed to capture deformations caused by excessive shoulder movements such as flexing and elevating, which occur as compensatory patterns during fatigue in bicep curl exercise. \textcolor{blue}{(C)} Display of the exemplar data from the subject performing weight-free bicep curls, standard bicep curls, and fatigued bicep curls. The derivation of the data, \(R_{norm}\) is discussed in Eq. (\ref{eq:RNorm}) }
    \label{fig:SensorOverview}	
\end{figure*}

\section{Development of the Proposed Wearable Shoulder Patch} 
\label{sec:development}

In this section, we detail the development of a new strain-sensor-based shoulder patch. First, we outline the hardware architecture of the sensor system. We then describe the algorithms developed for Real-Time and Post Hoc fatigue detection. An overview of the sensor system developed is shown in Fig. \ref{fig:SensorOverview}
\subsection{Hardware Development}

The strain sensor is fabricated using carbon nanofibers (CNFs) with Pyrograf-III, grade PR-24-XT-HHT reinforced platinum (Pt) thin films, as reported by our team's previous work \cite{peng2021carbon}. Specifically, the sensor demonstrates a gauge factor of approximately 417, a linear response rate of 97\%, a wide sensing range up to 121\% of strain, and excellent durability, maintaining stable performance over 2000 cycles of stretching and releasing. These features make it well-suited for detecting fatigue-induced compensatory movements. This work further adapts this sensor architecture to develop a soft wearable patch.

\subsubsection{Fabrication of Strain Sensor}

Briefly, CNFs are dispersed in isopropyl alcohol (IPA, Chem-Supply, 1 mg ml-1) with 2 wt\% Polyvinylpyrrolidone (PVP, Sigma, Mw \(\sim\) 40,000) as a dispersing agent. The mixture is probe-sonicated for one hour to prepare the conductive ink. Simultaneously, polydimethylsiloxane (PDMS, Sylgard 184) is fabricated by mixing the PDMS base and curing agent, degassing under vacuum, and molding.  The ink mixture is then spray-coated on the PDMS thin substrate using a spray gun (Blackridge airbrush kit, needle diameter 0.2 mm, pressure 15-20 PSI) from a distance of 10 cm and then dried in an air oven at $60\si{\celsius}$ for an hour. 

\begin{figure}[h!]	
    \centering

    \includegraphics[width=0.46\textwidth]{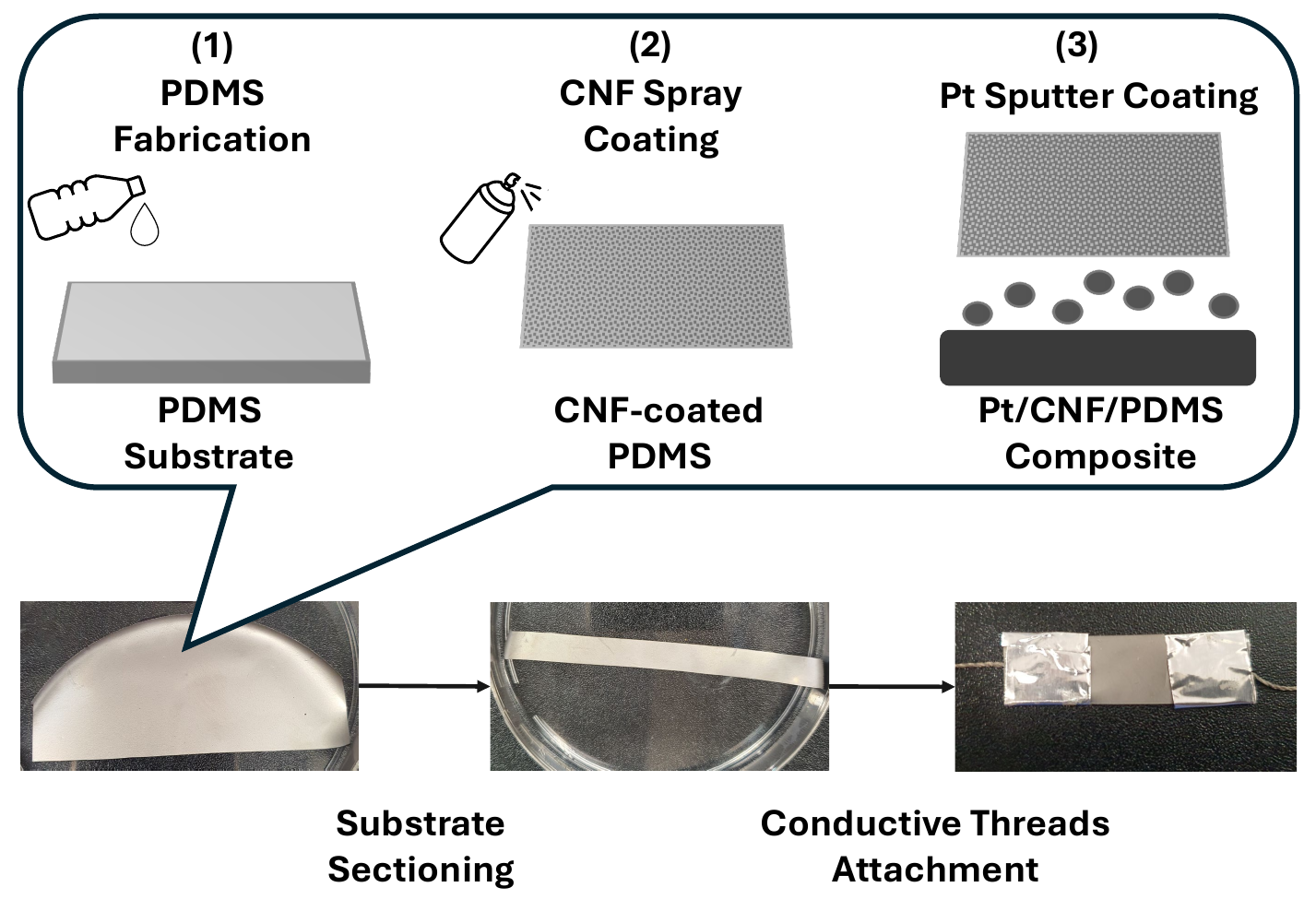}
    \caption{The manufacturing process of the strain sensor. \textcolor{blue}{(1)} Fabricating the PDMS substrates. \textcolor{blue}{(2)} Spray coating the CNF ink onto the PDMS substrates. \textcolor{blue}{(3)} Sputter coating the Pt on the CNF/PDMS substrates.}
    \label{fig:SensorManufacture}	
\end{figure}

Next, a 50\(\mu\)m layer of Pt coating is sputter-coated onto the surface of the CNF/PDMS substrates. Finally, the Pt/CNFs/PDMS is cut into 1 cm wide and 3 cm long strips with conductive threads attached to both ends using silver paste (OK-SPI, SPI Supplies) and encapsulated with aluminum foil tape. A comprehensive illustration of the sensor manufacturing process is provided in Fig. \ref{fig:SensorManufacture}.

\subsubsection{Wearable Design}
\begin{figure}[h!]	
    \centering

    \includegraphics[width=0.46\textwidth]{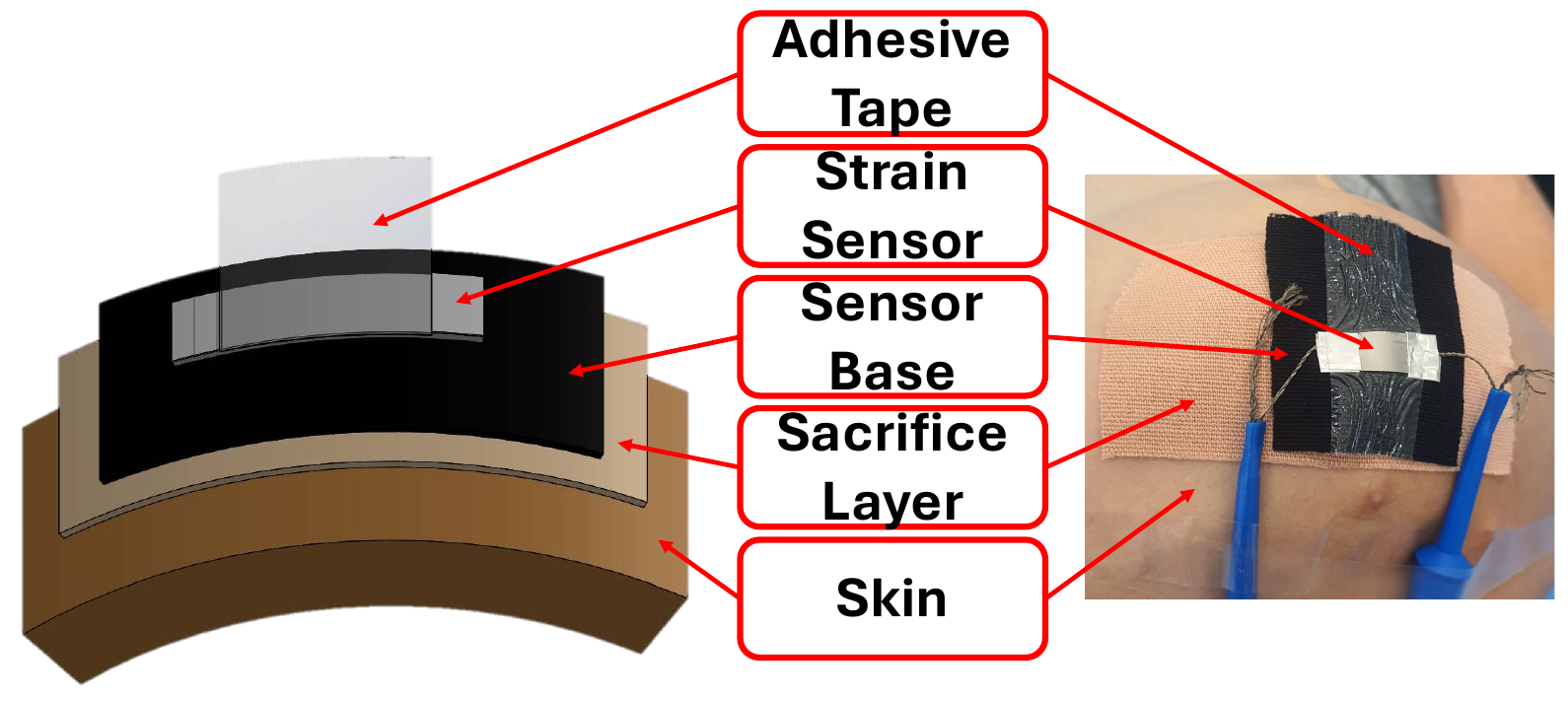}
    \caption{The detailed breakdown of the sensor component and its placement during use.}
    \label{fig:breakdown}	
\end{figure}

After fabrication, the sensor is mounted on kinesiology tape, which serves as the sensor's base, where transparent adhesive polyurethane (PU) tape is used for attachment. To ensure the sensor reusability, a sacrifice layer of the kinesiology tape first adheres the sensor to the targeted muscle or joint where the sensor is directly attached to this layer. The sacrifice layer is replaced after each use to maintain hygiene and ensure consistent sensor performance. The exploded view of the complete setup shows the skin and all components in Fig. \ref{fig:breakdown}.

\subsubsection{Sensor Interface Setup}
\begin{figure}[h!]	
    \centering
    \includegraphics[width=0.46\textwidth,keepaspectratio]{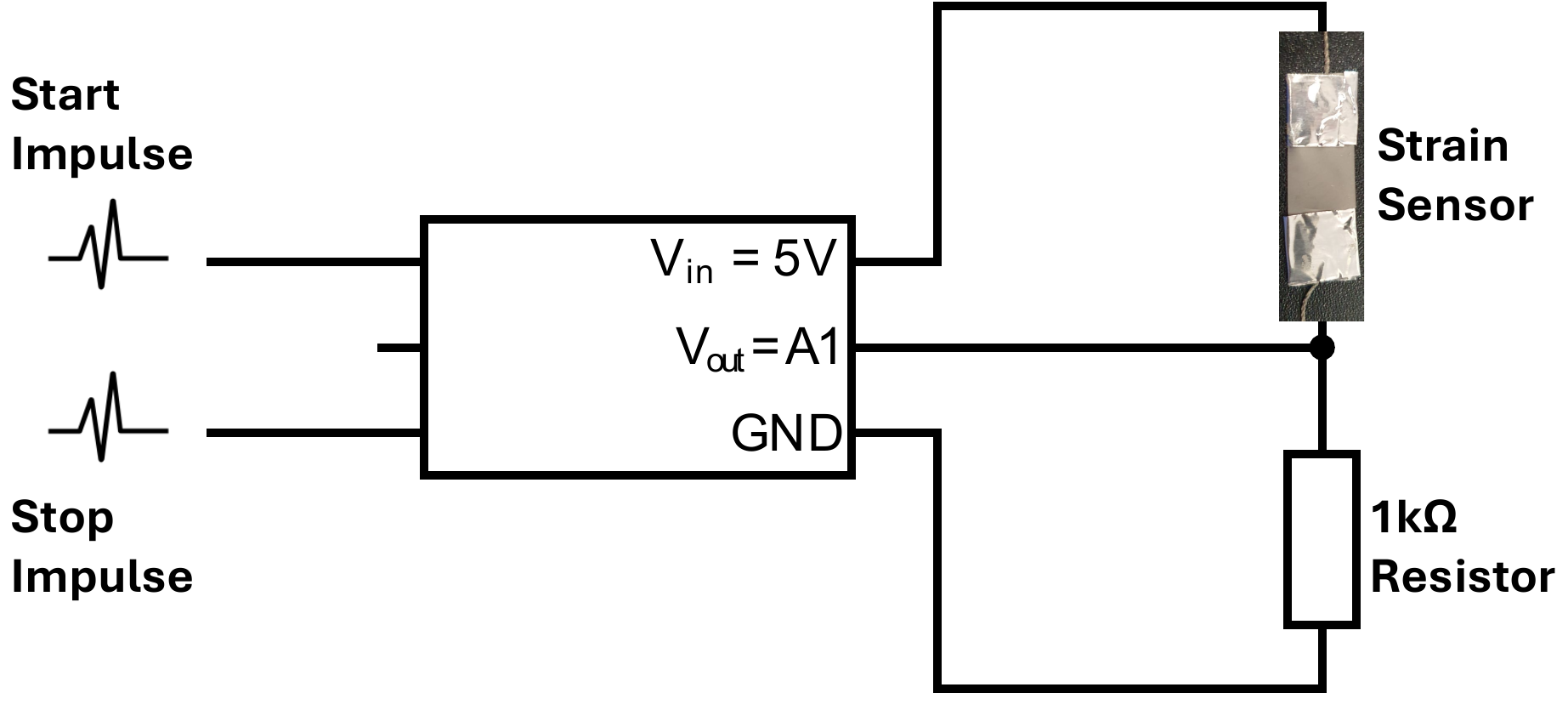}
    \caption{The functional block diagram of the wearable strain sensor circuit.}
    \label{fig:Schematic}	
\end{figure}

This section details the setup of the wearable strain sensor's interface. The sensor's resistance changes based on the applied strain. To interface the sensor, a voltage divider circuit is configured using a resistor with \(R=1\) k\(\Omega\), and an Arduino UNO as shown in Fig. \ref{fig:Schematic}. The sensor's conductive threads are mechanically secured to Arduino using a parrot hook. This setup allows the sensor to be integrated into the voltage divider, with the output voltage, \(V_{out}\) connected to the Arduino’s analog input pin, A1. In addition, the current setup takes in a start impulse and a stop impulse from the synchronization module to allow data collection. The resistance of the strain sensor at each sample, \(R_{raw}\) is calculated using the following equation:

\begin{equation}
    R_{raw} = R* (\frac{V_{in}}{V_{out}}-1) 
    \label{eq:VoltageDivider}
\end{equation}
where \(V_{in}\) represents the input voltage from Arduino and is typically 5V.
\subsubsection{Cost Analysis}
The primary components of the sensor including the CNFs, Pt, and PDMS, incur a material cost of approximately \(\$\)0.05 USD per unit. Additional materials that are needed to assemble the sensor such as the silver paste, aluminum foil, kinesiology tape, and conductive wire, raise the total to approximately \(\$\)0.74 USD per unit. All the materials may require replacement if they degrade over time. Meanwhile, the electronics hardware, such as the Arduino UNO, parrot hook, resistor, and breadboard, which is generally not subject to frequent replacement, adds a deployment cost of about \(\$\)34.77 USD per setup.
\subsection{Software Development}
Prior to the software development, preliminary experiments, as will be described in Sec. \ref{subsec:Ex3}, were conducted to determine the sensor's suitability to detect fatigue and to determine the key changes in the sensor signal associated with fatigue. These experiments concluded that placing the sensor on the AC joint is optimal for capturing fatigue-induced compensation during bicep curl exercises. Furthermore, the strain sensor effectively reflects the patient's fatigue condition during exercise. We observed that as fatigue increases, the strain signal exhibits a visible increment in the cycle amplitude and greater signal variability. 
Based on this observation, we propose two algorithms: 1) Real-Time fatigue detection and 2) Post Hoc fatigue detection. The former allows the detection of whether the patient is experiencing fatigue while exercising, while the latter enables the retrospective analysis of the patient's rehabilitation states.

\subsubsection{Strain Sensor Filtering \& Processing}
The Arduino is programmed to calculate and output the resistance value obtained from the strain sensor using Eq. (\ref{eq:VoltageDivider}). To reduce the sensor's noise, a median filter is applied to the signal to reduce the noise (we used MATLAB's built-in function “medfilt” in our implementation). Following the median filtering, the processed signal is normalized by dividing the filtered signal by the average signal value recorded during the static phase of the experiment using the following equation:
\begin{equation}
    R_{norm} = \frac{R_{raw} - \overline{R}_{static}}{\overline{R}_{static}}
    \label{eq:RNorm}
\end{equation}
where \(\overline{R}_{static}\) denotes the average strain sensor signal's value while the subjects remain stationary.

\subsubsection{Real-Time Detection} 
\begin{figure}[h!]
    \centering

    \includegraphics[width =0.49\textwidth,keepaspectratio]{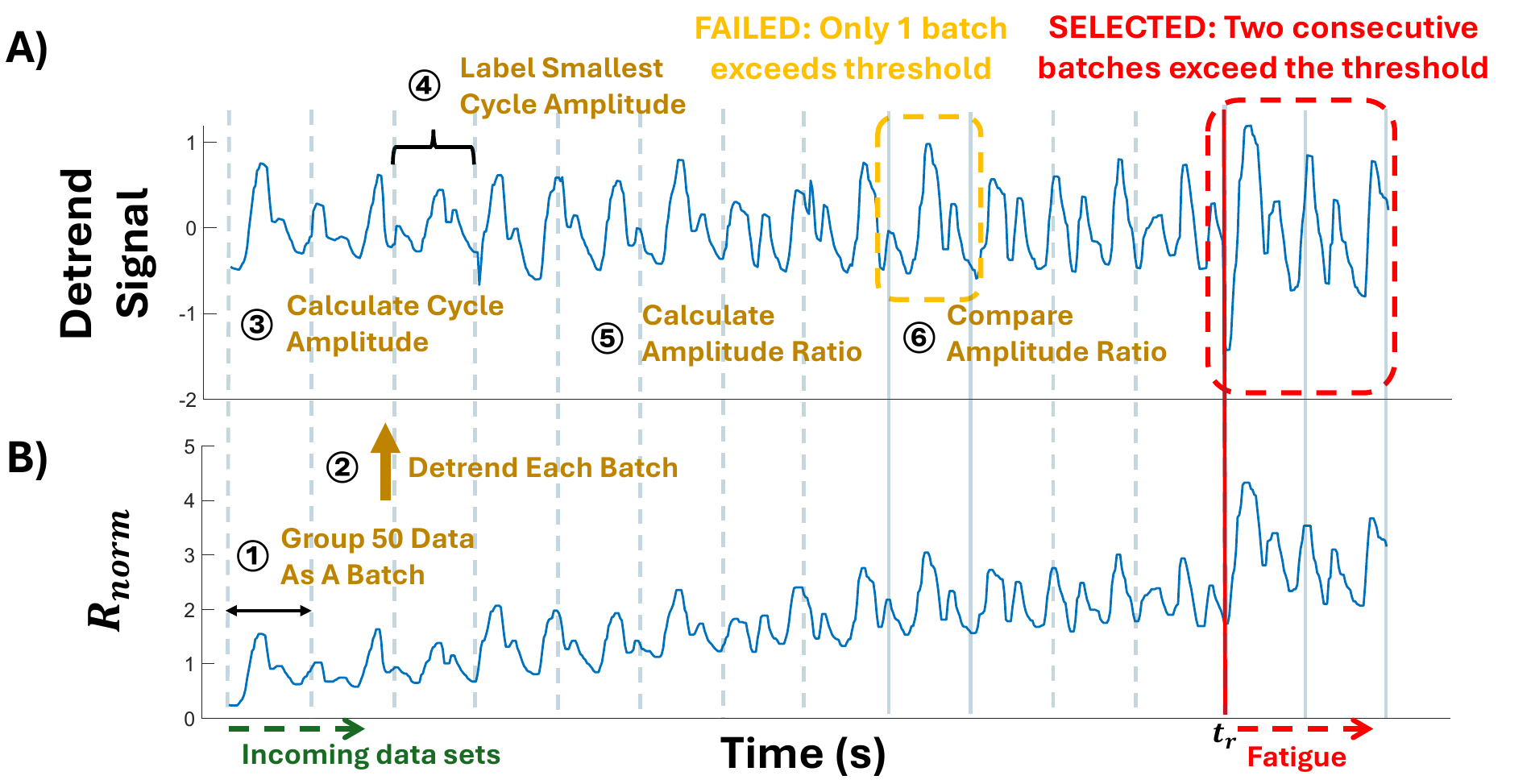}
    \caption{Real Time Strain Signal Processing. \textcolor{blue}{(A)} Detrended Strain Signal. \textcolor{blue}{(B)} Normalized Strain Resistance. The cycle amplitude of each signal batch is used to determine the fatigue time.}
    \label{fig:Online}	
\end{figure}

In the Real-Time mode, we aim to develop an algorithm that can identify when a subject enters the fatigue stage, along with the progress of the exercise. To this end, we propose a sliding window-based algorithm outlined in Algorithm \ref{alg:realTime} and the steps illustrated in Fig. \ref{fig:Online}. 

First, the preprocessed sensor signals \(R_{norm}\) are collected in batches of 50 data points for approximately 2s. This configuration is chosen due to an empirical observation that a bicep curl cycle takes approximately 2 seconds (step \textcircled{1}). Each batch is then detrended (step \textcircled{2}; we used MATLAB's “detrend” function in our implementation). Next, the peaks and troughs are identified in the detrended data, and the cycle amplitude is calculated (step \textcircled{3}). As each batch is processed, the minimum cycle amplitude is recorded and used as a reference for comparison (step \textcircled{4}). A fatigue threshold \(\tau\) of 3.5 is established by considering the linear relationship between the shoulder kinematics and the strain sensor that will be discussed in Sec. \ref{subsec:Ex2} and the increment of the shoulder elevation ROM between standard bicep curls and fatigued bicep curls \cite{chua2024analysis}. For each batch, the ratio of the current cycle amplitude to the reference cycle amplitude is calculated (step \textcircled{5}). If this ratio exceeds the predefined threshold for two or more consecutive batches, the start time of the first batch in the sequence is recorded as the fatigue time (step \textcircled{6}), denoted as \(t_{r}\). If the condition in step \textcircled{6} is not satisfied before the subject completes the exercise, \(t_r\) is set to be the end time of the signal, indicating the fatigue is not detected as the fatigue-induced compensation may not occur.
\begin{algorithm}
\caption{Real-Time Fatigue Detection} \label{alg:realTime}
\begin{algorithmic}
\STATE \textbf{Input:} Strain Sensor Signal \(R_{raw}\), Average Static Strain Sensor Signal \(\overline{R}_{static}\), Batch Size \(n\), Fatigue Threshold \(\tau\)
\STATE \textbf{Output:} Fatigue Timestamp \(t_r\), Fatigue Flag \(fat\)
\STATE \textbf{Initialization:} Fatigue Flag \(fat \gets \) \texttt{false}, Smallest Cycle Amplitude: \(Amp \gets \infty\), Consecutive Batch: \(numBatch \gets 0 \), Fatigue Timestamp: \(t_r \gets 0 \)
\WHILE{$R_{raw}$ is present}
    \STATE \(R_{norm} \gets\) Filter and preprocess \(R_{raw}\)
    \STATE Group \(R_{norm}\) into batch of size \(n\) \COMMENT{step \textcircled{1}}
    \FOR{each batch}
        \STATE \(R_{detrend} \gets\)Detrend each batch \COMMENT{step \textcircled{2}}
        \STATE \(currAmp \gets\) Identify peaks and troughs in \(R_{detrend}\) and calculate the cycle amplitude \COMMENT{step \textcircled{3}}
        \IF{\(currAmp < Amp\)}
            \STATE \(Amp \gets currAmp\) \COMMENT{step \textcircled{4}}
        \ENDIF
        \STATE \(ampRatio \gets \) Calculate amplitude ratio: \(currAmp \div Amp\) \COMMENT{step \textcircled{5}}
        \IF{\(ampRatio < \tau\)} 
            \STATE \(numBatch \gets numBatch + 1\) \COMMENT{step \textcircled{6}}
            \STATE \(t_r \gets \) Start time of current batch
               \IF{\(numBatch > 1\)} 
                    \STATE \(fat \gets \) \texttt{true}\\  
                    \textbf{break}
                \ENDIF
        \ELSE
            \STATE \(numBatch \gets 0\)
            \STATE \(t_r \gets \) 0
        \ENDIF
    \ENDFOR
\ENDWHILE
\RETURN \(t_r, fat\)

\end{algorithmic}
\end{algorithm}

\subsubsection{Post Hoc Detection}
In the Post Hoc mode, the full length of the preprocessed strain sensor signal \(R_{norm}\) is available. Compared with the Real-Time mode, this allows us to fully analyze the cycle amplitude of all bicep curls as well as the variability of the strain sensor signal. 

\begin{figure}[t!]	
    \centering

    \includegraphics[width =0.49\textwidth,keepaspectratio]{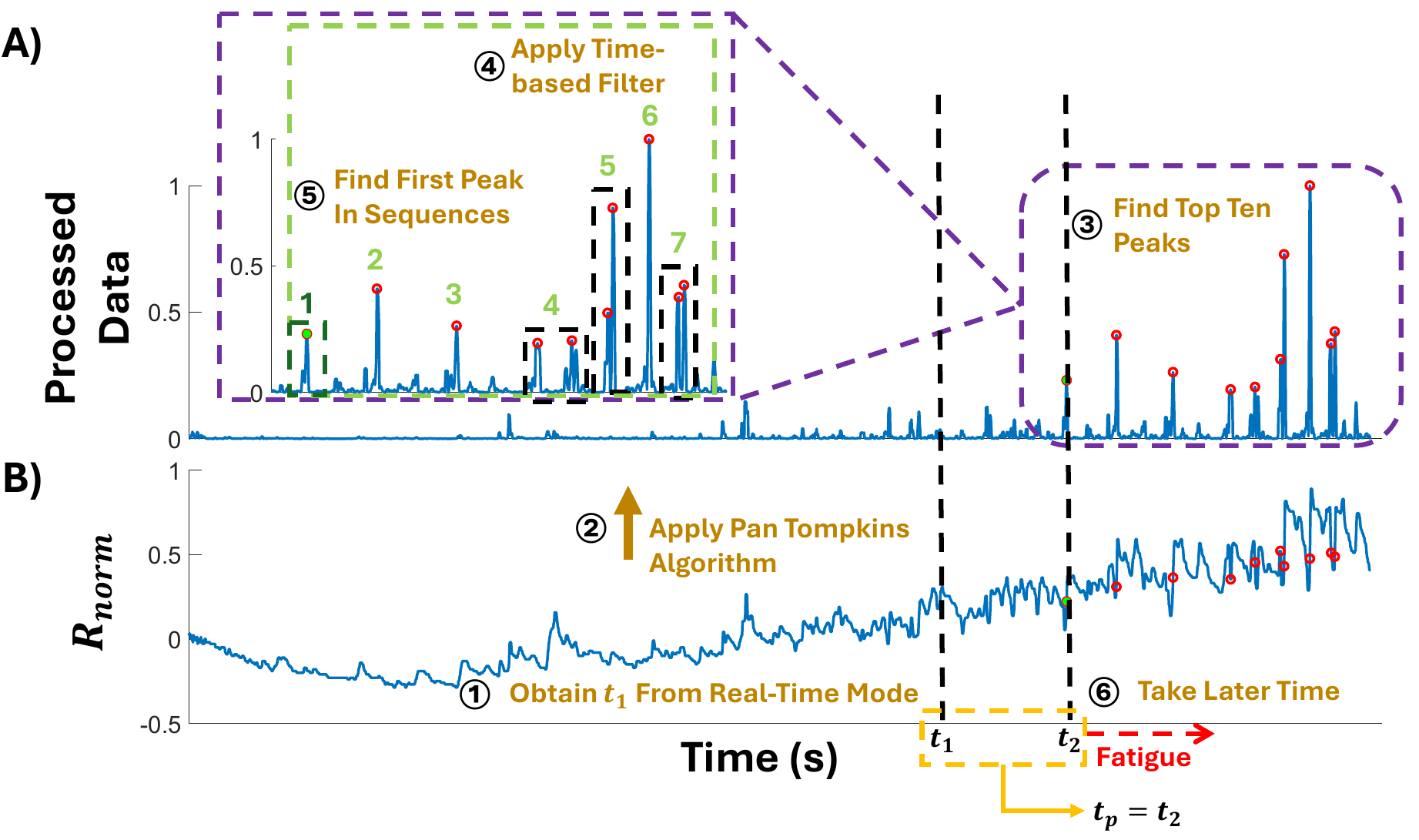}
    \caption{Post Hoc Strain Signal Processing. \textcolor{blue}{(A)} Post Pan-Tompskin Strain Data. The black box groups the peak groups from the same bicep curl, and the light green box with the number shows the consecutive peaks and their index. The dark green box shows the selected peak using the proposed algorithm. \textcolor{blue}{(B)} Normalized Strain Resistance. The time interval of the region with the highest signal variability is used to determine the fatigue time.
    }
    \label{fig:Offline}	
\end{figure}

The algorithm is outlined in Algorithm \ref{alg:postHoc} and the steps are illustrated in Fig. \ref{fig:Offline}. First, the same methodology described in the Real-Time mode is applied to obtain a timestamp \(t_1\) (step \textcircled{1}). Next, the Pan-Tompkins algorithm is applied to the normalized signal (step \textcircled{2}) to enhance the signal's gradient. The top ten peaks representing the signal region with the highest variability of the processed data are then selected (step \textcircled{3}). A time-based filter is applied to ensure that each detected peak is separated between 2 and 8 seconds (step \textcircled{4}), confirming that each peak corresponds to a distinct bicep curl cycle. Neighboring peaks that are separated by less than 2 seconds are considered as the same bicep curl cycle and will be treated as a single peak (e.g., peaks 4 \& 5). Peaks more than 8 seconds apart are considered signal noise and removed. 
The time of the first peak in the remaining continuous peaks (peak 1 in the illustrated case) is marked as the point of fatigue (step \textcircled{5}), denoted as \(t_{2}\). After obtaining both fatigue time onset, the later of \(t_1\) and \(t_2 \) is considered as the more accurate fatigue time onset, denoted as \(t_p\) (step \textcircled{6}).

\begin{algorithm}
\caption{Post Hoc Fatigue Detection} \label{alg:postHoc}
\begin{algorithmic}

\STATE \textbf{Input:} Strain Sensor Signal \(R_{raw}\), Average Static Strain Sensor Signal \(\overline{R}_{static}\),  Batch Size \(n\), Fatigue Threshold \(\tau\)
\STATE \textbf{Output:} Fatigue Timestamp \(t_p\), Fatigue Flag \(fat\)
\STATE \textbf{Initialization:} Fatigue Timestamp: \(t_p \gets 0 \)
\STATE \(t_1, fat \gets \) Compute Real Time Fatigue Detection :RealTime(\(R_{raw}\),\(\overline{R}_{static}\),\(n\),\(\tau\)) \COMMENT{step \textcircled{1}}
\STATE \(R_{norm} \gets\) Filter and preprocess \(R_{raw}\)
\STATE \(R_{pan} \gets\) Apply Pan Tompkins Algorithm on \(R_{norm}\) \COMMENT{step \textcircled{2}}
\STATE \(peaks \gets\) Identify top ten peaks in \(R_{pan}\) \COMMENT{step \textcircled{3}}
\STATE \(filteredPeaks \gets\) Apply time-based filter to combine or remove noises in \(peaks\), \COMMENT{step \textcircled{4}}
\STATE \(t_2 \gets\) Output the time of the first peak in \(filteredPeaks\) \COMMENT{step \textcircled{5}}
\STATE \(t_p \gets\) max(\(t_1,t_2\)) \COMMENT{step \textcircled{6}}

\RETURN \(t_p, fat\)

\end{algorithmic}
\end{algorithm}

\section{Experiment and Validation}
\label{sec:Ex&Valid}
In this section, we first describe the experimental setup and discuss the protocols employed in this study, including the bicep curl and fatigue protocols. Next, we outline the experiments conducted to determine the suitability of the sensors for detecting fatigue and to identify changes in the sensor signal associated with fatigue. This is followed by a detailed description of the subject study. Finally, we relate the findings from our previous study on fatigue during bicep curl to the proposed strain sensor and outline methods used to process data from different sensor systems to validate the sensor's efficiency.

\subsection{Experiment Setup}
\label{subsec:ExSetup}
\begin{figure}[th!]    
        \centering
\includegraphics[width=0.46\textwidth,keepaspectratio]{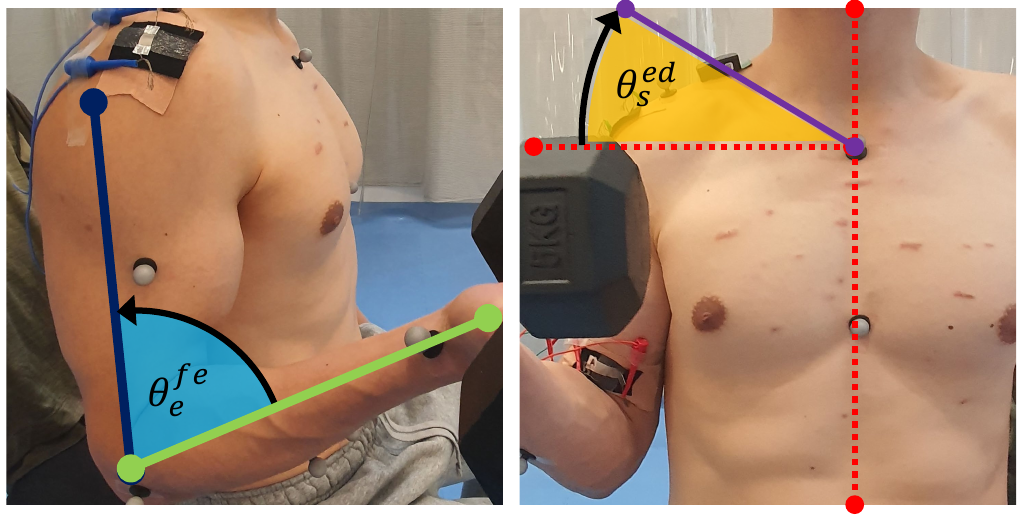}
    \caption{Two joint angles during bicep curl in this study.  The red dotted lines represent the references; the purple line links the clavicle and the acromioclavicular joint; the blue and green lines depict the upper arm limb and the lower arm limb, respectively. The arrows indicate the positive directions.}
    \label{fig:JointDef}
\end{figure}

\begin{figure*}[t!]	
    \centering
    \includegraphics[width=0.9\textwidth,keepaspectratio]{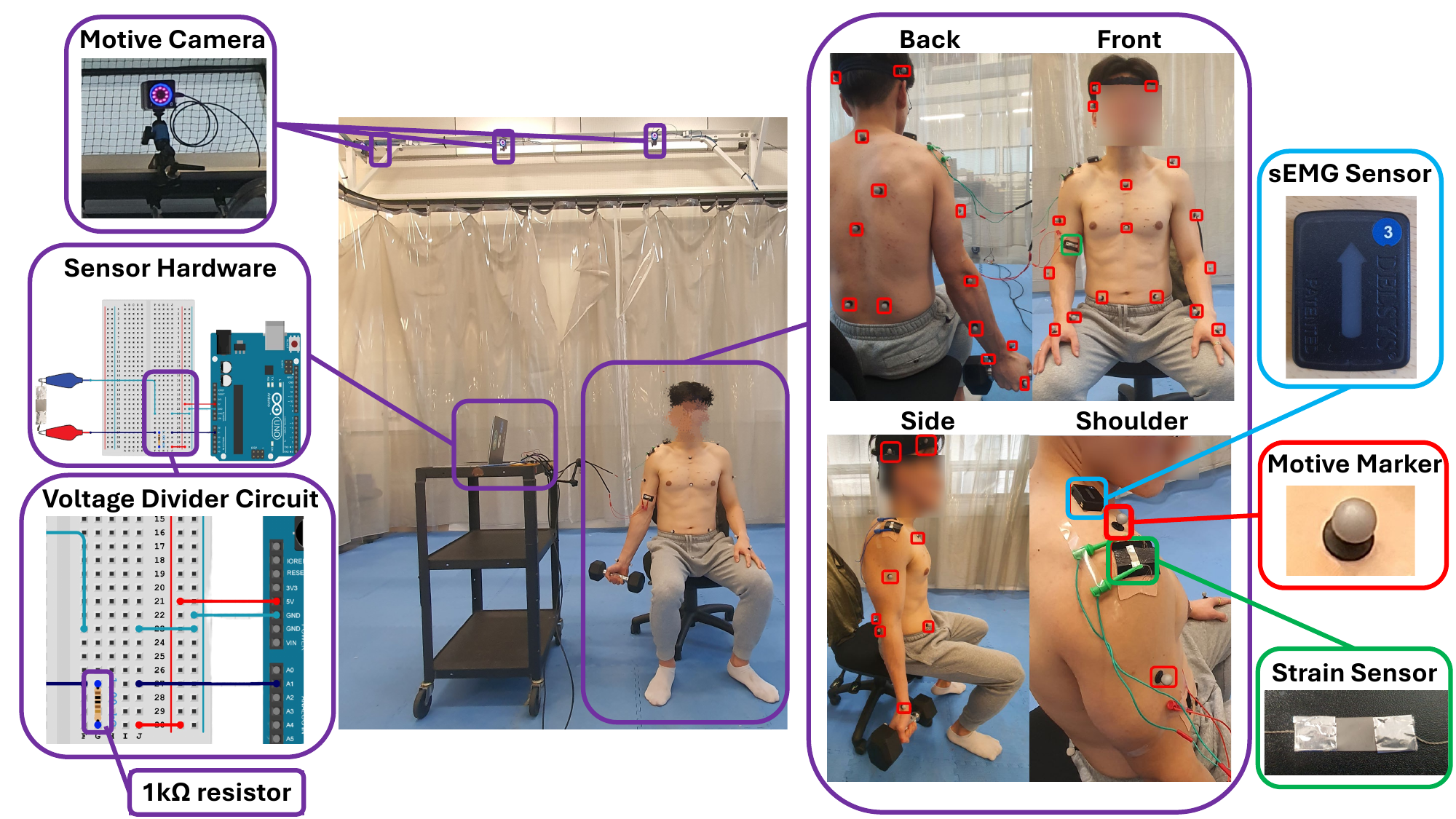}
    \caption{The experiment setup of the study and preliminary experiments; The apparatus used in this study and the sensor placement.}
    \label{fig:Experiment Setup}	
\end{figure*}

To conduct this study, we employed a system designed to capture detailed biomechanical and physiological data, incorporating sEMG sensors, a motion capture system, and our proposed strain sensor. The experiments were conducted at the Mechatronics Research Lab, School of Mechanical \& Manufacturing Engineering, UNSW Australia. The data acquisition involved ten motion capture cameras (PrimeX 13, OptiTrack, US) sampling at 100 Hz, one sEMG sensor (Trigno Avanti Sensor, Delsys, US) sampling at 1000 Hz, and one Arduino UNO sampling at 25 Hz. Twenty-seven reflective markers were placed on the subjects based on the Conventional Upper Model \cite{UpperLimbModelMotive} authored by Optitrack Motion System to capture the subjects' upper body joint kinematics. The system's software outputs the quaternion set of the skeleton joint angles, including the targeted \textit{elbow flexion-extension} \(\theta_{e}^{fe}\), and \textit{shoulder elevation-depression} \(\theta_{s}^{ed}\), whose definitions are shown in Fig. \ref{fig:JointDef}. 
The sEMG sensor was placed on the UT muscle based on the Seniam Guide \cite{Seniam}, and the strain sensor was affixed to the AC joint. To ensure the synchronization of data points across different devices, a synchronization module (eSync2, Optitrack, US) and a trigger module (Delsys Trigno Trigger Module, Delsys, US) were used to synchronize the data collection across the motion capture cameras, sEMG sensor, and Arduino UNO. A comprehensive illustration of the full experimental setup is provided in Fig \ref{fig:Experiment Setup}.

\subsection{Experiment Protocols}
There are different variations of bicep curls, and a universally accepted definition of fatigue has not been established. To ensure consistency throughout the experiment, we have specified the following protocols as reference.
\subsubsection{Bicep Curl Protocol}
To ensure consistency, the standard bicep curl is chosen, and all subjects should receive prior training. Subjects begin by supinating their right elbow joint to orient the palm upward. Then, they flex their elbow while maintaining the upper arm close to the trunk until the wrist reaches shoulder height. Then, they extend their elbow back to the starting position. In this context, a complete bicep curl cycle is defined from the initiation of elbow flexion to the full return of the elbow joint to its initial posture.

\subsubsection{Subject's Fatigue Protocol}
To assess the subject's perceived exertion and fatigue during the study, we employ the Borg CR-10 Scale \cite{borg1982psychophysical}. During the experiment, subjects self-evaluate based on their RPE, with a level of seven as the threshold of fatigue. Fatigue is defined either by the inability to complete a bicep curl cycle or verbal confirmation of the inability to continue. At that point, the subject's perceived onset of fatigue, marked by timestamp, \(t_s\), is recorded.

\subsection{Preliminary Experiments}
Four preliminary experiments were conducted to evaluate the suitability of the strain sensor for detecting changes in kinematics, sEMG signals, and fatigue. It is important to note that the sensor placements in these preliminary experiments were different from those described in Section \ref{subsec:ExSetup}, which was specific to the subject study. 
\begin{figure}[h!]	
        \centering
    \includegraphics[width=0.46\textwidth,keepaspectratio]{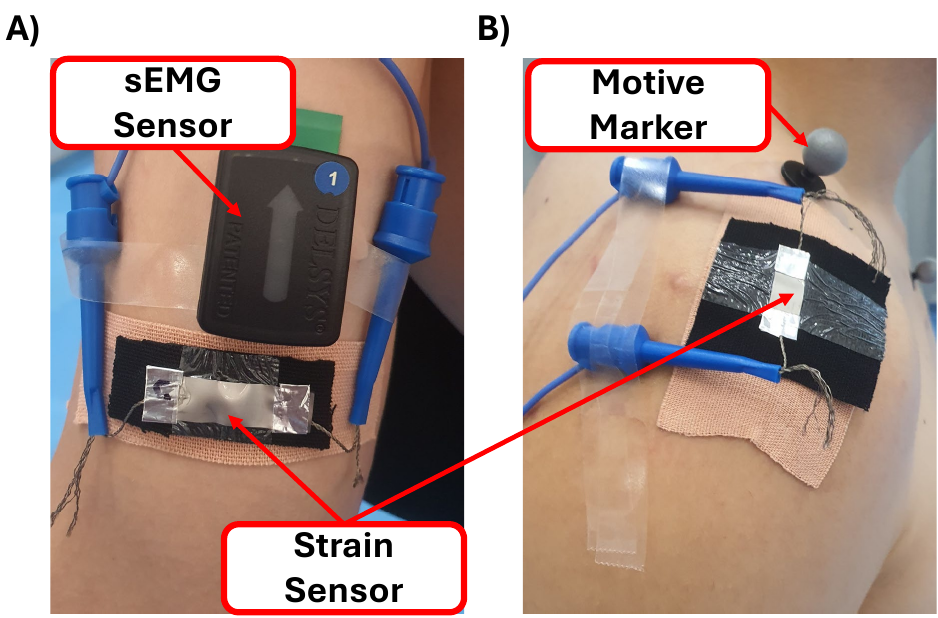}
    \caption{The sensor placement for the preliminary experiments. \textcolor{blue}{(A)} The sEMG sensor and the strain sensor are placed on the BIC muscle. \textcolor{blue}{(B)} The motive marker and the strain sensor are placed on the shoulder joint.}
    \label{fig:PreEx}	
\end{figure}
\subsubsection{Experiment 1: Static-Noise and Power Consumption Assessment of Strain Sensor}
\begin{figure}[h!]	
        \centering
    \includegraphics[keepaspectratio,width=0.46\textwidth]{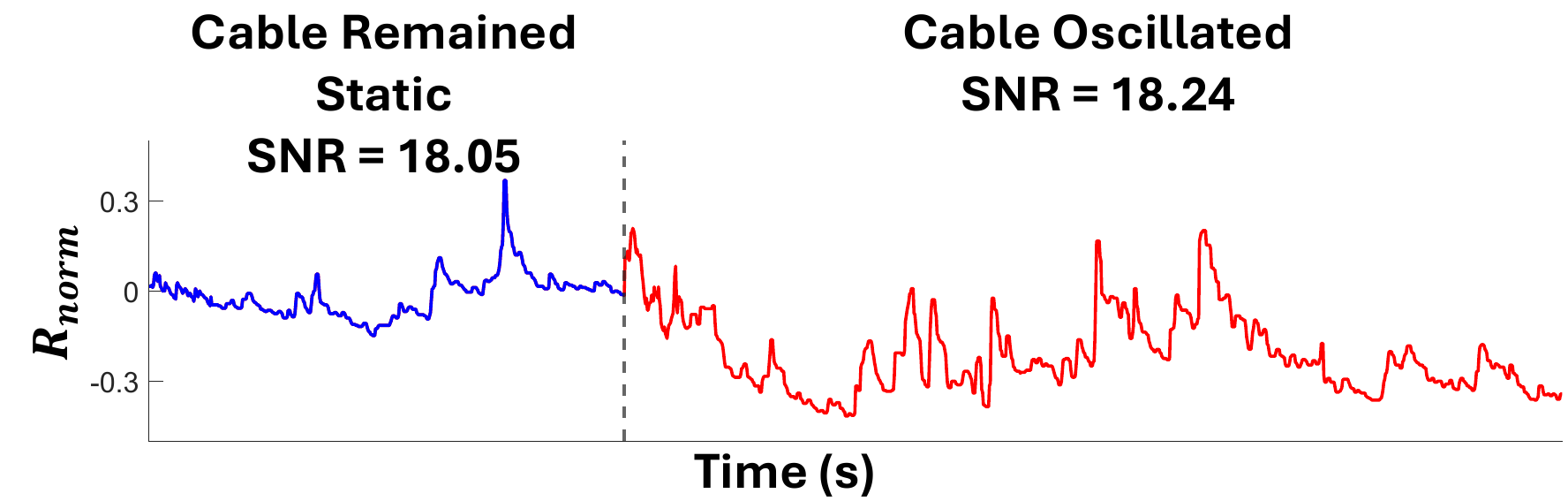}
    \caption{Comparison of the Normalized Strain Resistance under two test conditions. The blue graph represents the condition where the cable is left undisturbed. The red graph represents the condition where the cable is vigorously oscillated. The signal-to-noise ratio remains similar, indicating that cable motion does not degrade the signal quality.}
    \label{fig:SensorNoise}	
\end{figure}

This experiment aimed to quantify the sensor’s baseline static noises and additional noise when the system is disturbed, recognizing the signal acquired from human skin, a highly deformable surface and the signal quality is constrained by the acquisition system. The experimental setup used was identical to that illustrated in Fig. \ref{fig:Experiment Setup}, excluding the sEMG sensor and Motive markers. The outputs from the strain sensor were recorded while the subject remained static under two conditions, where the interconnecting cables remained undisturbed and when the interconnecting cables were deliberately jostled to emulate the worst-case handling. As illustrated in Fig. \ref{fig:SensorNoise}, our application reported a signal-to-noise ratio ranging from 18.0 dB to 28.1 dB, indicating a consistently clean signal. The result also indicated no reduction in the signal-to-noise ratio when the cable is vigorously manipulated. This preliminary study concluded that as long as the parrot hook is secured on the subject's shoulder, it does not affect the functionality of the strain sensor. 

This experiment also quantifies the nominal power consumption of the circuit. The strain sensor forms a voltage divider circuit with a 1k \(\Omega\) reference resistor, \((R)\) and it is excited from a regulated 5.0 V from Arduino, \((V_{in})\). At static, the nominal strain resistance, \((R_{strain})\) is recorded at approximately 700 \(\Omega\). Using Equation \ref{eq:Power}, the circuit is dissipating 14.7 mW.
\begin{equation}
    P = \frac{V_{in}^2}{R+R_{strain}} 
    \label{eq:Power}
\end{equation}

\subsubsection{Experiment 2: Comparison of Changes in Strain Resistance Against sEMG}
This experiment aimed to determine whether the strain sensor could accurately capture changes corresponding to those in the sEMG sensor output. The overall experimental setup used was similar to the one illustrated in Fig. \ref{fig:Experiment Setup} with slight modifications on the sensor placement of the strain sensor and sEMG sensor on the bicep brachii (BIC) muscle as shown in Fig. \ref{fig:PreEx}A. The outputs from both sensors were recorded while the subject performed five bicep curls under weight-free and standard conditions.

\begin{figure}[h!]	
        \centering
\includegraphics[keepaspectratio,width=0.46\textwidth]{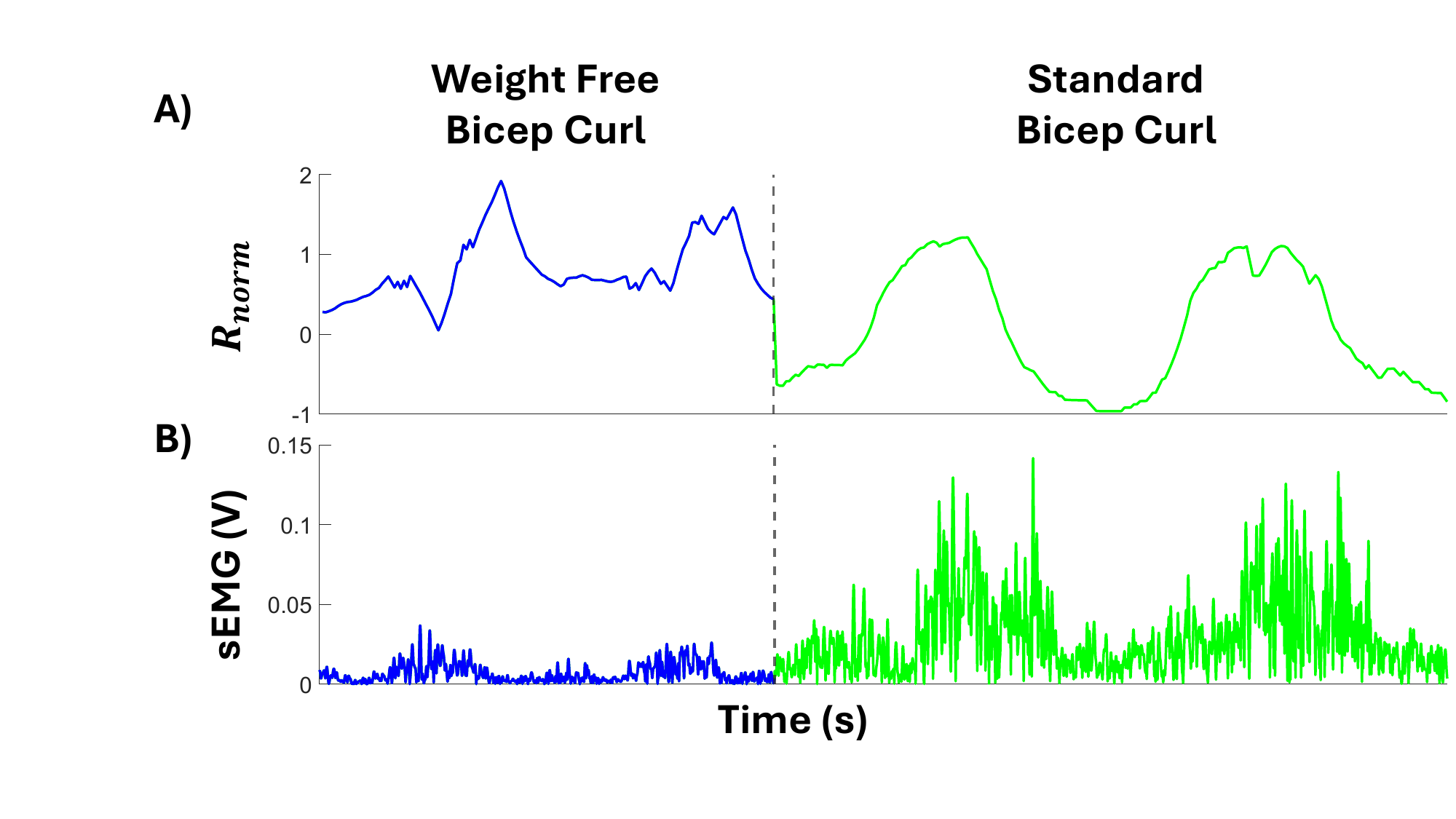}
    \caption{Comparison of sEMG Signals and Strain Resistance Across Different Exercise Intensities. The blue graph represents the signal collected under a weight-free bicep curl, while the green graph represents the signal collected under a standard bicep curl. \textcolor{blue}{(A)} Normalized Strain Resistance. \textcolor{blue}{(B)} The RMS of sEMG signal.}
    \label{fig:SensorEMG}	
\end{figure}

The result is presented in Fig. \ref{fig:SensorEMG}. The figure shows two \(R_{norm}\) and sEMG RMS pair under two conditions: weight-free bicep curl and standard bicep curl. Our study observed that the maximum amplitude of the sEMG RMS from the BIC muscle is approximately 0.05 V during weight-free bicep curl, increasing to 0.15 V during standard bicep curl. In contrast, the corresponding changes in the resistance measured by the strain sensor were minimal. Conducting Cross Correlation Analysis, the signal under the weight-free condition has a positive correlation of 0.58 with a lag of 1.43 seconds while the signals under the standard condition have a positive correlation of 0.49 with a lead of 0.3 seconds.

These findings suggest that our strain sensor may not be well-suited for detecting fatigue via changes in sEMG amplitude, as these variations are scarcely detectable. The lack of significant resistance changes may be due to insufficient skin deformation during exercises, which limits the sensor's ability to detect subtle changes. 
This interpretation aligns with findings reported in the literature on the challenges of measuring skin deformation with strain sensors in dynamic muscle activity scenarios \cite{alvarez2022towards}. 

\subsubsection{Experiment 3: Comparison of Changes in Strain Resistance Against Shoulder Elevation Kinematics}
This experiment aimed to determine whether the strain sensor could accurately capture the changes corresponding to those in the shoulder elevation trajectory. The overall experimental setup used was similar to that in Fig. \ref{fig:Experiment Setup} with slight modifications on the sensor placement of the strain sensor on the AC joint as shown in Fig. \ref{fig:PreEx}B. The experiment records the subject's shoulder kinematics and the strain sensor signal while conducting shoulder elevation with varying amplitudes.

\begin{figure}[h!]	
        \centering
\includegraphics[width=0.46\textwidth,keepaspectratio]{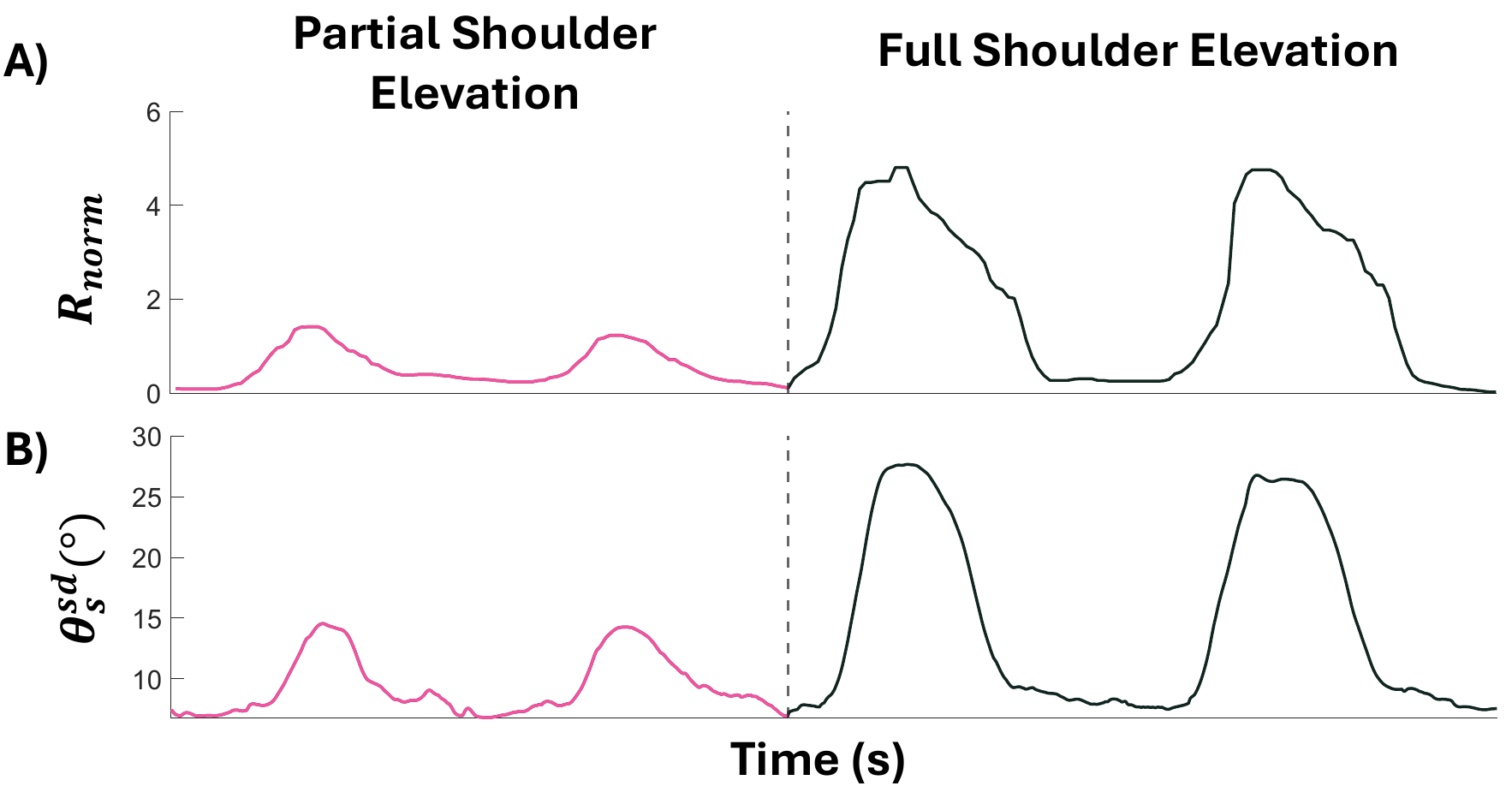}
    \caption{Comparison of Shoulder Elevation Angle and Strain Resistance Across Different Shoulder Elevation Range. The pink graph represents the signal collected under partial shoulder elevation, while the black graph represents the signal collected under full shoulder elevation. \textcolor{blue}{(A)} Normalized Strain Resistance. \textcolor{blue}{(B)} The shoulder elevation angle.}
    \label{fig:SensorKin}	
\end{figure}

The result is presented in Fig. \ref{fig:SensorKin}. The figure shows \(R_{norm}\) and the shoulder elevation trajectory during partial shoulder elevation and full shoulder elevation. Our results demonstrate that the resistance changes of our strain sensor correlate with shoulder joint movements. When the shoulder elevation ROM increased from 7 to 20 degrees, the maximum normalized resistance increased from 1.3 to 4.8, exhibiting a similar trend to the joint kinematics trajectory. The incremental ratio between the strain sensor signal and shoulder elevation ROM is approximately 1.4, indicating a proportional relationship between joint movement and skin deformation captured by the sensor. Conducting Cross Correlation Analysis, the signal in both stages achieved a high correlation of 0.92 and 0.95 with a lag of 0.01 seconds.

In summary, the strain sensor signal exhibited a moderate correlation with sEMG measurements (r = 0.49), indicating a reasonable association between muscle activation and strain during the standard bicep curl. In contrast, a strong correlation was observed between the strain signal and shoulder kinematics (r = 0.95), suggesting that the sensor is highly sensitive to compensatory joint movements. These findings support the use of the strain sensor as an effective surrogate for detecting fatigue-induced compensation, particularly through observable kinematic changes.

This observation aligns with the previous studies that applied similar sensor technology to detect changes in human joints, akin to applications seen in hand sign detection \cite{peng2021carbon}. Thus, we conclude that the outputs of the strain sensor change proportionally with the change in shoulder elevation angle, making it an optimal tool for detecting fatigue by monitoring the shoulder elevation ROM.
\label{subsec:Ex2}

\subsubsection{Experiment 4: Changes in Strain Resistance During Continuous Bicep Curls}
This experiment aimed to determine whether the strain sensor can reflect changes in the subjects' condition during bicep curl and to identify which sensor characteristics are suitable for detecting the onset of muscle fatigue. The experimental setup used was identical to that illustrated in Fig. \ref{fig:Experiment Setup}, excluding the sEMG sensor and Motive markers. In a single continuous session, subjects performed a series of bicep curls, starting with weight-free conditions and transitioning into standard bicep curls until reaching fatigue. Finally, subjects performed bicep curls in their fatigue state. The sensor measurements were recorded, and the collected data were analyzed to identify signal patterns or metrics indicating the transitions between stages.

\begin{figure}[h!]	
    \centering

    \includegraphics[keepaspectratio,width =0.49\textwidth]{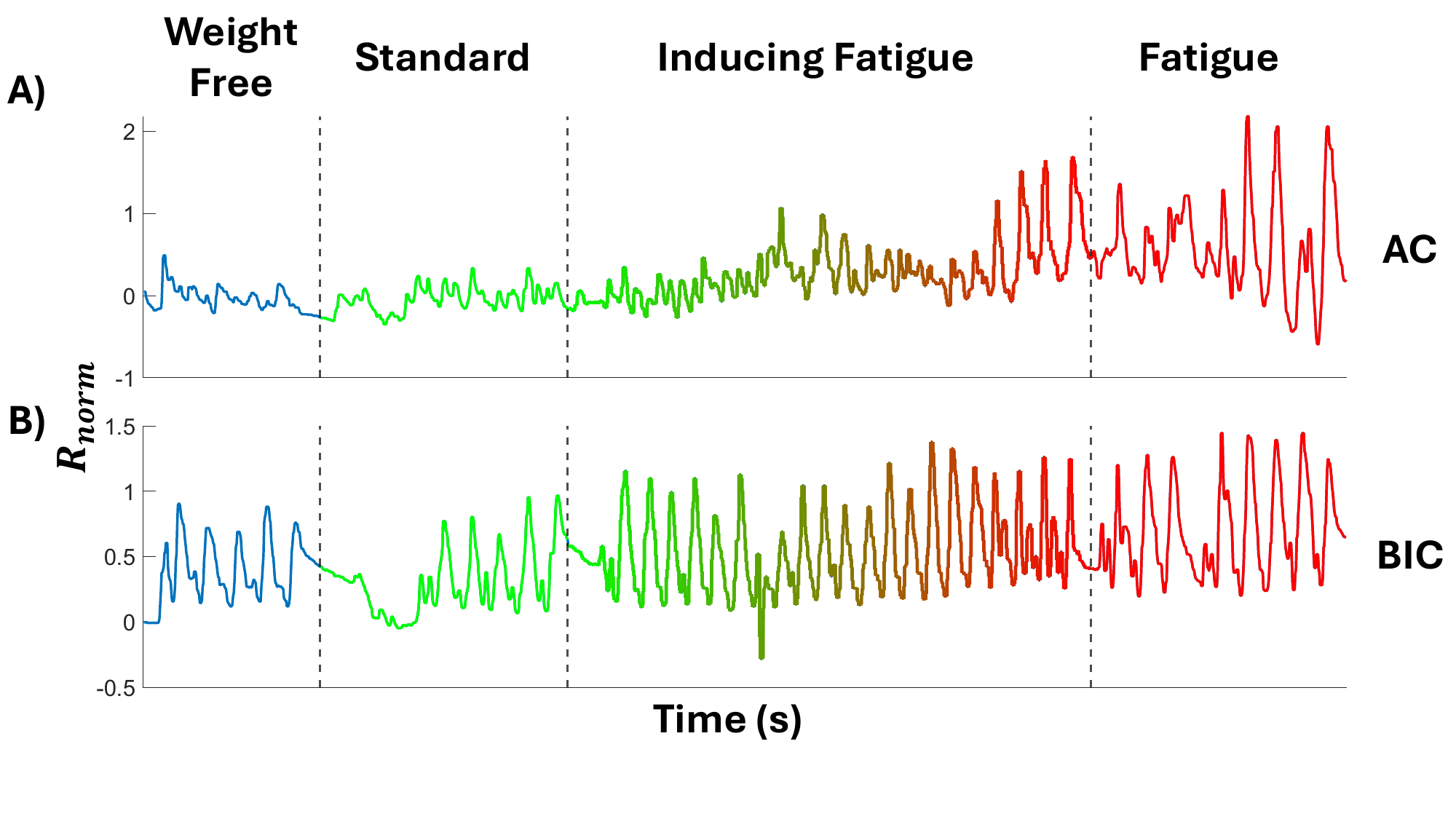}
    \caption{Changes in Normalized Strain Resistance During Continuous Bicep Curls. The blue graph represents the signal collected under weight-free bicep curl, the green graph represents the signal collected under standard bicep curl, and the red graph represents the signal collected under fatigue bicep curl. \textcolor{blue}{(A)} Normalized Strain Resistance of AC Joint. \textcolor{blue}{(B)} Normalized Strain Resistance of BIC Muscle.}
    \label{fig:ex3}	
\end{figure}

The data collected are presented in Fig. \ref{fig:ex3}, showing \(R_{norm}\) from both the AC joint and the BIC muscle. The data collected during weight-free bicep curl is shown in the blue graph, while the data collected during standard bicep curl is shown in green, transitioning gradually into red to represent the data collected during fatigue bicep curl.

It is observed that each trough-peak-trough sequence in the normalized resistance data collected from the sensor on the BIC muscle represents one complete bicep curl cycle, and its cycle amplitude does not show substantial changes throughout the experiment. Conversely, for the sensor on the AC joint, the cycle amplitude of the sensor resistance changes significantly across different stages of the experiment. During weight-free and standard bicep curls, the amplitude fluctuates around 0.5 and gradually increases to approximately 2 as the subject continues to perform bicep curls until the fatigue stage. Furthermore, the strain signal from the AC joint exhibits increased strain and greater variability. From a biomechanical perspective, as fatigue sets in during the bicep curl exercise, the subject’s control over maintaining shoulder stability decreases, leading to compensatory movements using the shoulder to complete the exercise. This biomechanical compensation explains the observed higher variability and increased amplitude in the sensor resistance from the AC joint.
These observations led to the decision to place the sensor on the AC joint for effective fatigue detection, emphasizing that the cycle amplitude and signal variability as critical criteria for identifying fatigue.

\label{subsec:Ex3}

\subsection{Subject Study}
A subject study was conducted to validate the proposed sensor's efficiency. The UNSW Human Research Ethics Committee approved the experiment under the reference number HC220573. 

\subsubsection{Subjects}
Thirteen male subjects were recruited for this study, with ages ranging from 18 to 35, body weights ranging from 65kg to 100kg, and heights ranging from 165cm to 185cm. None of them reported any physical injuries, disabilities, or history of stroke. The subjects declared engaging in exercise for one to more than seven hours per week. A summary of the subjects' demographics is tabulated in Table \ref{table:Subjects}. 
All subjects provided written informed consent before their inclusion in the study. Participation in the study was voluntary, with the option for subjects to withdraw their participation at any time before publication of the results.
\begin{table}[h!]
    \centering
    \caption{Subjects Demographics}
    \begin{tabular}{ccccc}
        \hline
        Subject & \makecell{Age \\(year)} & \makecell{Height \\ (cm)} & \makecell{Weight \\(kg)} & \makecell{Exercise duration \\ per week (hour)}\\
        \hline
        
        1 & 24-29 & 175-180 & 90-100 & 5-7\\  
        2 & 30-35 & 175-180 & 80-90 & 3-5\\  
        3 & 24-29 & 175-180 & 70-80 & 7+\\  
        4 & 18-23 & 170-175 & 70-80 & 7+\\  
        5 & 18-23 & 175-180 & 80-90 & 5-7\\  
        6 & 24-29 & 180-185 & 80-90 & 5-7\\  
        7 & 18-23 & 170-175 & 70-80 & 3-5\\  
        8 & 18-23 & 175-180 & 70-80 & 1-3\\  
        9 & 30-35 & 165-170 & 60-70 & 3-5\\  
        10 & 24-29 & 175-180 & 60-70 & 5-7\\  
        11 & 24-29 & 170-175 & 80-90 & 3-5\\  
        12 & 24-29 & 170-175 & 60-70 & 1-3\\
        13 & 30-35 & 170-175 & 60-70 & 3-5\\  \hline
    \end{tabular}
    \label{table:Subjects}
\end{table}

\subsubsection{Experiment Procedure}
On the experiment day, the authors checked the sensors to ensure they were generating stable data points before the subject's arrival. Upon arrival, the authors explained the bicep curl and fatigue protocols to the subjects and verified they met all the inclusion criteria. Once confirmed, the strain sensor, sEMG sensor, and motion capture markers were appropriately positioned.

The subject sat on a stool and remained stationary while baseline static data were collected for three seconds. Subsequently, the subjects performed five bicep curls. After a brief pause, they continued performing bicep curls until reaching a state of fatigue. Finally, after another brief pause, they performed five post-fatigue bicep curls.

\subsection{Evaluation Metrics}
\begin{figure}[h!]	
    \includegraphics[keepaspectratio,width=0.49\textwidth]{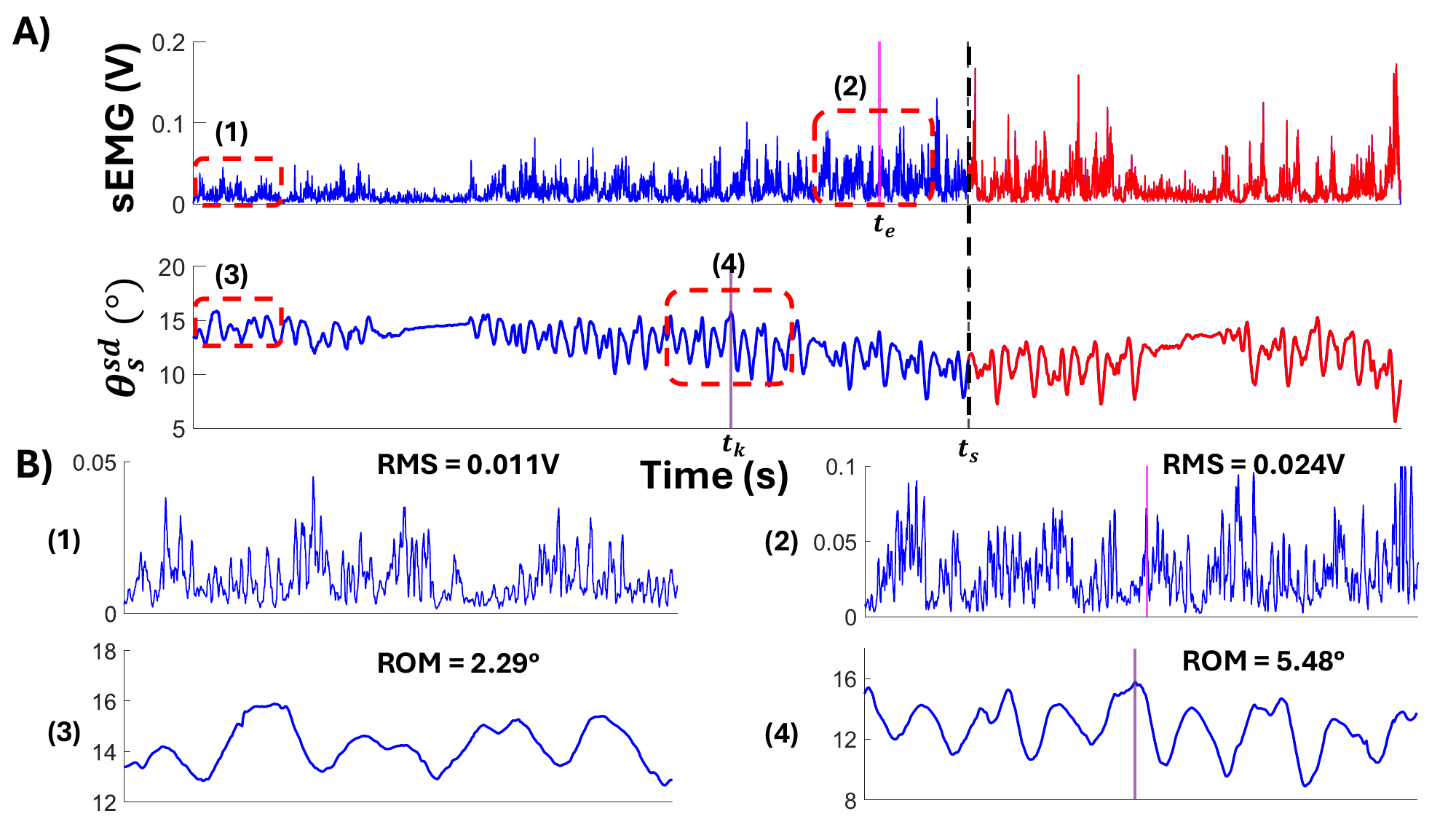}
    \caption{Kinematics and sEMG Fatigue Detection. The blue plot represents the signal collected before the subject's perceived fatigue time, and the red plot represents the signal collected after. \(t_s,t_e,t_k\) represents the fatigue time nominated by the subjects, the fatigue time determined by the sEMG signal, and the fatigue time determined by the shoulder elevation angle, respectively. \textcolor{blue}{A)} The RMS amplitude of the UT sEMG and the shoulder elevation angle. The black dotted line represents the subject's perceived fatigue time, separating the strain sensor signal collected under standard bicep curl and fatigue bicep curl. \textcolor{blue}{B)} The highlights of the selected data. \textcolor{blue}{(1)} The amplitude of sEMG RMS when the subject conducts standard bicep curl. \textcolor{blue}{(2)} The amplitude of sEMG RMS when the subject conducts fatigue bicep curl. The pink line represents the fatigue time determined via sEMG. \textcolor{blue}{(3)} The ROM of shoulder joint elevation during standard bicep curl. \textcolor{blue}{(4)} The ROM of shoulder joint elevation during standard bicep curl. The purple line represents the fatigue time determined via shoulder elevation.}
    \label{fig:EMGKin}	
\end{figure}

To objectively identify the onset of fatigue, the RMS of the sEMG from the UT muscle and the joint kinematics of shoulder elevation are analyzed. The section below details the processing methods used to derive \(t_e\) from the sEMG and \(t_k\) from the shoulder joint kinematics data. Fig. \ref{fig:EMGKin} illustrates obtaining the fatigue time using sEMG and joint kinematics data. The figure shows the RMS sEMG and the shoulder elevation angle collected during the subject study.

\subsubsection{sEMG of UT}
The sEMG data collected undergoes processing, including a median filter followed by a bandpass filter within the frequency range of 10 Hz to 150 Hz to eliminate noise interference. The RMS amplitude, denoted as \(E_{RMS}\), is then calculated to quantify the muscle activation level:
\begin{equation}
	E_{RMS} = \sqrt{\frac{1}{n} \sum_{i} E_{i}^2}
\end{equation}
where \(E_i\) represents the sEMG signals value at each sample and \(n\) represents the number of samples in the RMS window. 

A sliding window is applied to the RMS sEMG, with a threshold set at a 127\% increase, based on the RMS levels observed during standard bicep curls to detect significant deviations. When the average RMS within a window exceeds this threshold, the start time of that window is noted as the fatigue time \(t_e\) (pink line), as indicated by the sEMG signal.

\subsubsection{Shoulder Elevation Kinematics}

The quaternion sets from Motive are transformed using YZY Euler angle sequences to calculate the plane of elevation, elevation angle, and internal-external rotation angle. Similarly, the attitude of the forearm relative to the upper arm is analyzed using XZY Euler angle sequences, enabling the extraction of flexion-extension and internal-external rotation angles \cite{cutti2005soft}.

The joint kinematics trajectory is analyzed to identify peaks and troughs. The cycle amplitude between each consecutive peak and trough was calculated. An increase of 150\% over the cycle amplitude observed during standard bicep curls is used as a threshold. When the cycle amplitude for three consecutive trough-peak-trough sequences exceeds this threshold, the timestamp of the first peak in this sequence is recorded as the onset of fatigue \(t_k\) (purple line), as indicated by joint kinematics.

\section{Result and Discussion}
In this section, we present and discuss the fatigue times obtained from sEMG and shoulder elevation angles. We then analyze the fatigue times obtained from the developed strain sensor system using both Real-Time and Post Hoc methods. The result obtained is tabulated in Table \ref{table:Fatigue}.
\label{sec:result}

\subsection{Obtaining Fatigue Timestamp}
\subsubsection{sEMG} 
In our study, fatigue time was successfully captured based on the sEMG signal from 8 out of 13 subjects. The time differences between \(t_e\) and \(t_s\) were calculated. The average difference between these two metrics was 15.35 seconds, with a standard deviation of 19.50 seconds, ranging from 1.86 seconds to 32.2 seconds, excluding data from subjects where fatigue was undetected. For the undetected subjects, the maximal differences in RMS computed between the reference and other windows was approximately 1.5.

The variability in the sEMG signal from the UT muscle does not consistently indicate fatigue using the threshold from previous studies, particularly in subjects with high weekly exercise durations (over 5 hours), except for Subject 4. This observation aligns with Jurasz's findings on exploring the neuromuscular fatigue responses during cycling exercise \cite{jurasz2022neuromuscular}. Their research suggests that the sEMG amplitude of the recreationally active subjects group increased significantly compared to the athlete group when the workload increased. In addition, subjects with higher physical fitness often have better techniques and control over movements and often manage fatigue more efficiently, focusing effort more effectively on targeted limbs and joints, hence compensating less.

\subsubsection{Kinematics}
In our study, fatigue time was successfully captured based on shoulder kinematics from 12 out of 13 subjects. The time differences between \(t_k\) and \(t_s\) was calculated. The average time difference was 21.2 seconds, with a standard deviation of 27.48 seconds, ranging from 2.69 seconds to 65.8 seconds.

For the subject where fatigue was not detected through shoulder kinematics, the maximum increment of the ROM between the standard bicep curl and other curls was approximately 62\%, which is significantly lower than the set threshold. From the previous study, the numerical changes between the standard and fatigue bicep curls ranged from 4 to 10 degrees. Given the relatively small size of the shoulder joint, with a clavicle length of 13 to 16 cm, these small numerical differences are likely to lead to magnified percentage changes.

Additionally, variations in detection times were observed among subjects. Subject 12 showed an early detection, and Subject 8 showed non-detectability, potentially due to their lower weekly exercise periods. For Subject 4, early detection was observed and believed to be influenced by their distinct exercise habits, which likely involve different types of muscle engagement compared to other subjects.

\subsection{Subject Study: Real-Time Detection}
The algorithm successfully captured the fatigue time for all subjects, and the resulting fatigue timestamps indicate a tendency for early detection. Specifically, the average absolute time difference from \(t_s\) was 21.65 seconds, with a standard deviation of 21.80 seconds, ranging from 2.1 seconds to 65.3 seconds. 

This Real-Time method identifies the point at which a subject's range of motion continuously exceeds a predefined threshold. Notably, subjects who reported exercising more than 7 hours per week exhibit highly similar detection results between the online and kinematics methods. This similarity underscores the ability of these subjects to maintain consistent and stable exercise forms throughout the experiment. Their robust physical conditioning likely contributes to more predictable and uniform sensor readings, which align closely with kinematic assessments. In addition, despite not exercising more than 7 hours per week, Subject 2 showed nearly identical results from both the Real-Time method and the kinematics-based assessments.

For other subjects, this level of similarity in trend was less pronounced. Variability among these individuals can be attributed to a range of factors, including diverse exercise habits and different physiological responses to fatigue. Less experienced exercisers may not consistently maintain proper form throughout their sessions, particularly as fatigue sets in. This inconsistency can lead to alternating patterns in their movements, such as muscle tremors, which can falsely influence the detected shoulder elevation ROM.

\subsection{Subject Study: Post Hoc Detection}
The efficacy of the proposed Post Hoc method for detecting fatigue was assessed using an algorithm that reliably captured fatigue time for all participants. The fatigue time was successfully captured for all subjects. Results indicated a tendency for late detection, with an average absolute time difference compared with \(t_s\) of 9.54 seconds, with a standard deviation of 7.70 seconds, ranging from 0.63 seconds to 19.4 seconds. 12 out of 13 subjects' fatigue time was detected from the Pan Tompskin algorithm method, while only subject 6 inherited the fatigue time from the Real-Time method. 

In contrast with the Real-Time method, which demonstrates higher accuracy for subjects with weekly exercise duration, the Post Hoc method does not prefer any specific exercise group. The nature of selecting the latter time significantly reduces the high error cases for some subjects (subjects 4, 5, and 11). Despite increasing the error for some subjects (subjects 1, 2, 3, 8, and 9), most of the difference remains between 0.26 seconds and 5.64 seconds, with one outlier of 10.66 seconds. 

\begin{table}[h!]
    \centering
    \caption{Fatigue Timestamp}
    \begin{tabular}{cccccc}
        \hline
         \multirow{4}{*}{Subject} &  \multirow{3}{*}{\makecell{Subject \\declared \\fatigue \\time \(t_s\)}}& \multicolumn{4}{c}{\makecell{Time differences \\ with subject's declaration (s)}} \\ \\ \cline{3-6}
           &  & \makecell{Kinematics\\ \(t_s-t_k\)} & \makecell{sEMG\\ \(t_s-t_e\)}  & \makecell{Real-Time\\ \(t_s-t_{r}\)} & \makecell{Post Hoc\\ \(t_s-t_{p}\)}\\   \hline

1                         & \multicolumn{1}{c|}{58.97}      & -20.43              & NA                    & -2.12                 & -12.79                \\
2                         & \multicolumn{1}{c|}{48.88}      & 11.97               & 1.85                  & 13.77                 & -19.42               \\
3                         & \multicolumn{1}{c|}{55.99}      & -10.24              & NA                    & -10.77                & -13.16               \\
4                         & \multicolumn{1}{c|}{101}        & 65.87               & 22.35                 & 65.31                 & -2.00                \\
5                         & \multicolumn{1}{c|}{71.2}       & 36.17               & NA                    & 25.77                 & -0.63                \\
6                         & \multicolumn{1}{c|}{58.82}      & 18.45               & NA                    & -4.59                 & -4.59                \\
7                         & \multicolumn{1}{c|}{68.57}      & -11.38              & -32.19                & 43.17                 & -16.84               \\
8                         & \multicolumn{1}{c|}{34.33}      & NA                  & -6.80                  & 5.22                  & -5.48                \\
9                         & \multicolumn{1}{c|}{50.68}      & -10.54              & -5.27                 & 2.60                  & -5.94                \\
10                        & \multicolumn{1}{c|}{51.96}      & -8.81               & NA                    & 26.19                 & -14.72               \\
11                        & \multicolumn{1}{c|}{75.46}      & 2.69                & -13.2                 & 41.24                 & -8.84                \\
12                        & \multicolumn{1}{c|}{70.79}      & 46.52               & 26.34                 & 24.40                 & 8.97                 \\
13                        & \multicolumn{1}{c|}{64.23}      & -12.00              & 14.83                 & 16.31                 & -10.68               \\ \hline
Avr 1                     & \multicolumn{1}{c|}{}           & 22.26              & 32.29                & 21.65                & 9.54                 \\
     & \multicolumn{1}{c|}{}           & (27.23)              & (33.41)                &  (21.80)                & (7.70)                   \\\hline
Avr 2                     &  \multicolumn{1}{c|} {}         & 21.20                & 15.35                & 21.65                 & 9.54                   \\
 &  \multicolumn{1}{c|} {}         & (27.48)                & (19.50)                & (21.80)                 & (7.70)                   \\ \hline

    \end{tabular}
     \begin{tablenotes}
      \item[a] NA: Not detected.
     \item[b] Avr 1: This is the average value and the standard deviation of the absolute time differences, including the non-detected data, substituting the non-detected data with 0.
     \item[c] Avr 2: This is the average value and the standard deviation of the absolute time differences, excluding the non-detected data.
     \item[d]  The values in Avr 1 and Avr 2 are reported as Mean (Standard Deviation). 

  \end{tablenotes}
    \label{table:Fatigue}
\end{table}

\subsection{Limitation}
The proposed system successfully realized fatigue detection during bicep curl exercises in Real-Time and Post Hoc settings. Despite these advantages, the following limitations are acknowledged, and improvements can be made in the following fields:
\subsubsection{Definition on Fatigue}
Defining and detecting fatigue objectively remains a challenge. Although the inability to complete a full-range bicep curl was consistently used as a metric, verifying the consistency of subjects' self-reported fatigue timestamps is problematic. This is crucial as these self-reports are the primary ground truth relied upon, and any inconsistency in the definition of the fatigue point could lead to variable findings.
\subsubsection{Movement Isolation}
Throughout the study, subjects were instructed to sit to isolate upper limb movements, which limited the engagement of other muscle groups like the core muscles. However, it remains challenging to conclude that the force exerted during the exercise was solely from the upper limb muscles.
\subsubsection{Exercise Frequency Factor}
Preliminary observations suggest a correlation between exercise frequency, intensity, and the consistency of fatigue detection. However, the current sample size is too small to draw definitive conclusions. Future studies should aim to recruit a larger cohort to validate these findings and explore the impact of different exercise regimes on fatigue detection accuracy.

\subsection{Home-based Application}
The sensor system is designed to facilitate a direct transition from laboratory settings to in-home use, ensuring maximum user-friendliness. Setup requires minimal effort and can be performed with assistance from a caregiver or family member without medical knowledge. The AC joint can be easily located by identifying the small protrusion on the shoulder. Once identified, the sensor application involves simply placing a layer of kinesiology tape as a sacrifice layer over this area, positioning the sensor on top, and firmly securing the connection wire on the back of the patient, thereby rendering the sensor system ready for use.  
\section{Conclusion}
\label{sec:Conclusion}
This study introduces a wearable shoulder patch with a single strain sensor for detecting fatigue during bicep curl exercises. We developed and tested two algorithms, Real-Time and Post Hoc, to analyze the data captured by our system. The study involved thirteen subjects, and the results from our sensor-based fatigue detection were systematically compared with subjects' self-reported fatigue times, kinematics-based detections, and sEMG-based fatigue times.

The findings demonstrate that this approach offers a cost-effective solution for identifying fatigue-induced muscle compensation during exercise. The results indicate that kinematics-based detection successfully captured the majority of the subjects' fatigue points. Based on this, we hypothesize that integrating kinematic sensors, such as IMU, could significantly enhance the accuracy and robustness of fatigue detection. IMUs would provide supplementary information on body dynamics, enriching the data obtained from traditional strain sensors and potentially improving the overall effectiveness of the wearable device in real-time fatigue assessment. Building on these promising results, our team is currently developing a more advanced wearable device that incorporates an IMU. This integration aims to further enhance the accuracy and reliability of fatigue detection by providing additional data on body dynamics.

In addition, our team is developing a complete wearable unit that integrates the sensor with a more user-friendly design, eliminating the need for an Arduino and long connector wires that currently limit movement and make the system impractical for home-based use. Future iterations will consider some commercial low-power microcontrollers, such as ARM Cortex M0, STM32L, and ESP32-S2, to enable a more compact and efficient solution for wearable applications. 

\bibliographystyle{IEEEtran}
\bibliography{ref.bib}

\end{document}